\documentclass{PoS}

\usepackage{epsfig}
\usepackage{latexsym}
\usepackage{amsmath,amssymb}
\usepackage{textcomp}

\title{Two-Baryon Correlation Functions in 2-flavour QCD}

\ShortTitle{Two-Baryon Correlation Functions in 2-flavour QCD}

\author{Anthony Francis$^1$, \speaker{Chuan Miao}$^1$, Thomas D. Rae$^2$\thanks{Supported by DFG grant HA4470/3-1}, Hartmut Wittig$^{1,2,3}$\\
        $^1$~Helmholtz-Institut Mainz, Johannes-Gutenberg Universit\"at Mainz, \\
        $^2$~Institut f\"ur Kernphsyik, Johannes-Gutenberg Universit\"at Mainz, \\
	$^3$~PRISMA Cluster of Excellence, Johannes-Gutenberg Universit\"at Mainz\\
        E-mail: \email{francis@kph.uni-mainz.de, chuan@kph.uni-mainz.de, thrae@kph.uni-mainz.de, wittig@kph.uni-mainz.de}}

\abstract{We present first results for two-baryon correlation functions, computed using
$N_f=2$ flavours of O($a$) improved Wilson quarks, with the aim
of explaining potential dibaryon bound states, specifically the H-dibaryon. 
In particular, we use a GEVP to isolate the groundstate using two-baryon (hyperon-hyperon) correlation functions $\big(\langle
C_{XY}(t)C_{XY}(0) \rangle$, where $XY=\Lambda\Lambda, \Sigma\Sigma, N\Xi,
\cdots\big)$, each of which has an overlap with the H-dibaryon. 
We employ  a `blocking' algorithm 
to handle the large number of contractions, which may easily be
extended to N-baryon correlation functions. We also comment on its application
to the analysis of single baryon masses ($n$, $\Lambda$, $\Xi$, $\cdots$).
This study is performed on an isotropic lattice with $m_\pi = 460$ MeV, $m_\pi
L = 4.7$ and $a = 0.063$ fm.}

\FullConference{31st International Symposium on Lattice Field Theory LATTICE 2013\\
                 July 29 - August 3, 2013\\
                 Mainz, Germany}

\begin{document}

\section{Introduction}

Predicting bound states and the interaction of a multi-baryon system remains
a difficult challenge in lattice QCD. In the strange sector, model-dependent theoretical
studies suggest the existence of a flavour singlet dibaryon state, the so-called
H-dibaryon \cite{Jaffe}. Recent dynamical lattice QCD calculations have reported hints of
such states in the SU(3) flavour limit \cite{NPL1,NPL2,HAL}. Unfortunately, due to the lack of a
study including the volume dependence, as well as controlled chiral and
continuum extrapolations, there is no definitive conclusion at this time. Here, we present an initial study of two-baryon correlation functions with the aim of explaining potential dibaryon bound states away from the SU(3)-flavour limit.

\section{Multi-baryon correlation functions on the lattice}

\begin{figure}[b!]
\begin{center}
\includegraphics[width=0.85\linewidth]{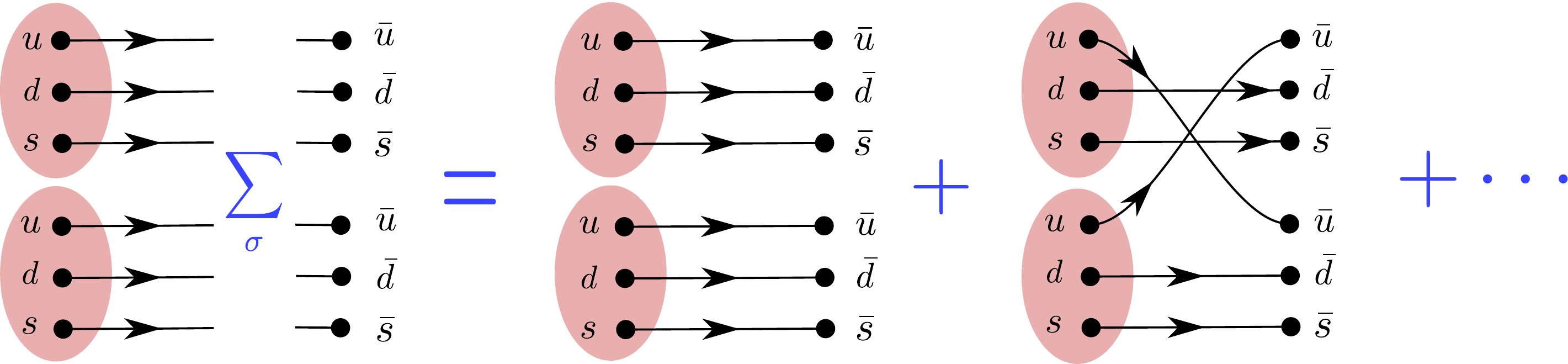}
\caption{A blocking algorithm simplifies the required contractions of the six-quark operator. The two individual baryons on the source side are pre-contracted into baryon blocks that can then be contracted using permutation matrices. For illustration purposes we show here the $\Lambda\Lambda-\Lambda\Lambda$ channel.}
\label{fig:blocking2}
\end{center}
\end{figure}

We analyze hyperon-hyperon states that have an overlap with the H-dibaryon. The lattice correlation functions we compute are
\begin{equation}
\label{eqn:corr}
\langle O_{X_1Y_1}(t)O_{X_2Y_2}(0) \rangle,~~~~~ \mathsf{where}~\  
X_iY_i = \Lambda\Lambda, \Sigma\Sigma, N\Xi.
\end{equation} 
Here we define the two-baryon interpolating operators to be
\begin{align}
\Lambda\Lambda(x) &= \epsilon^{ijk} \epsilon^{lmn} 
    (u^{i\mathsf{T}} C\gamma_5 d^j) (u^{l\mathsf{T}} C\gamma_5 d^m) (s^{k\mathsf{T}} C\gamma_5 s^n)\ ,\\
\Sigma\Sigma(x) &=\epsilon^{ijk} \epsilon^{lmn}  
    (u^{i\mathsf{T}} C\gamma_5 s^j) (d^{l\mathsf{T}} C\gamma_5 s^m) (u^{k\mathsf{T}} C\gamma_5 d^n)\ ,\\
N \Xi (x) &= \epsilon^{ijk} \epsilon^{lmn} 
    (u^{i\mathsf{T}}  C\gamma_5 d^j) (d^{l\mathsf{T}} C\gamma_5 s^m) (u^{k\mathsf{T}} C\gamma_5 s^n)\ .
\end{align}
Note that the source $X_2Y_2$ and sink $X_1Y_1$ operators in Eq.~\ref{eqn:corr} need not be the same, which in turn gives the following six independent correlation functions:
\begin{align}
X_1Y_1 - X_2Y_2=\langle O_{X_1Y_1}(t)O_{X_2Y_2}(0) \rangle=
\begin{pmatrix}
 \Lambda\Lambda -\Lambda\Lambda \quad & \Lambda\Lambda -\Sigma\Sigma \quad &\Lambda\Lambda-N\Xi \\
 &  \Sigma\Sigma -\Sigma\Sigma \quad  & \Sigma\Sigma -N\Xi\\
 &  & N\Xi  -N\Xi 
\end{pmatrix}\quad.
\label{eq:bbmatrix}
\end{align}
Results for all six correlation functions are shown in this study.
In principle, all of these correlation functions on the lattice have an overlap with the H-dibaryon ground state.  However, it is not possible a priori to determine which of the six possibilities has the greatest overlap. Additionally, computing all terms enables one to set up a generalized eigenvalue problem (GEVP) \cite{GEVP1,GEVP2,GEVP3} to determine the lowest lying  masses in the system.

\subsection{Blocking algorithm}

The number of contractions required for the multi-baryon correlation functions is equal to the factorials of their respective quark contents, $N_{contr}=N_{u}! N_{d}! N_{s}!$.  For each of the six correlation functions of Eq.~\ref{eq:bbmatrix} eight terms need to be contracted, whereas for the deuteron, for example, 36 contractions have to be handled. The factorial growth of the number of contraction terms requires more sophisticated methods on the algorithmic side of the calculation to solve this problem.
Here, we choose to implement a blocking procedure, similar to that proposed in \cite{Endres,Det} . The essential ingredient is pre-contracting the baryons on the source side of the correlation function. These baryon blocks can then be used to to contract the sink indices efficiently.
Taking, for example, the $\Lambda\Lambda-\Lambda\Lambda$ channel we have,
\begin{align}
\langle O_{\Lambda\Lambda}(t) O_{\Lambda\Lambda}(0) \rangle
&=(C\gamma_5)_{\alpha\beta}\sum_{\sigma_u, \sigma_d, \sigma_s}
f(\alpha, \xi^\prime_{\sigma_u(1)}, \xi^\prime_{\sigma_d(2)}, \xi^\prime_{\sigma_s(3)})
f(\beta,  \xi^\prime_{\sigma_u(4)}, \xi^\prime_{\sigma_d(5)}, \xi^\prime_{\sigma_s(6)})\\
&\times \epsilon_{c_1^\prime, c_2^\prime, c_3^\prime} \epsilon_{c_4^\prime, c_5^\prime, c_6^\prime}
(C\gamma_5)_{\alpha_2^\prime \alpha_3^\prime } (C\gamma_5)_{\alpha_5^\prime \alpha_6^\prime }
\end{align}
where the sum is over all possible contractions $\sigma_{u,d,s}$ and the blocks 
\begin{equation}
f(\alpha_1, \xi_1^\prime, \xi_2^\prime, \xi_3^\prime) =
\epsilon_{c_1, c_2, c_3} (C\Gamma_5)_{\alpha_2\alpha_3}
S_u(\xi_1,\xi_1^\prime) S_d(\xi_2,\xi_2^\prime) S_s(\xi_3,\xi_3^\prime)
\end{equation} 
are calculated beforehand. Here $\alpha$ is the open sink Dirac index, $\xi$/$\xi^\prime$ denote the
combined colour (Roman) and Dirac (Greek) indices at the sink/source, while $S_{u,d,s}$
is a propagator of flavour $u, d, s$, see Fig.~\ref{fig:blocking2} for an illustration. An advantage of this kind of algorithm is a straight-forward generalization from the hyperon-hyperon system studied here to multi-baryon systems with a mass number $A>2$ opening up the possibility to compute light nuclei.


\begin{figure}[t!]
\begin{center}
\includegraphics[width=\linewidth]{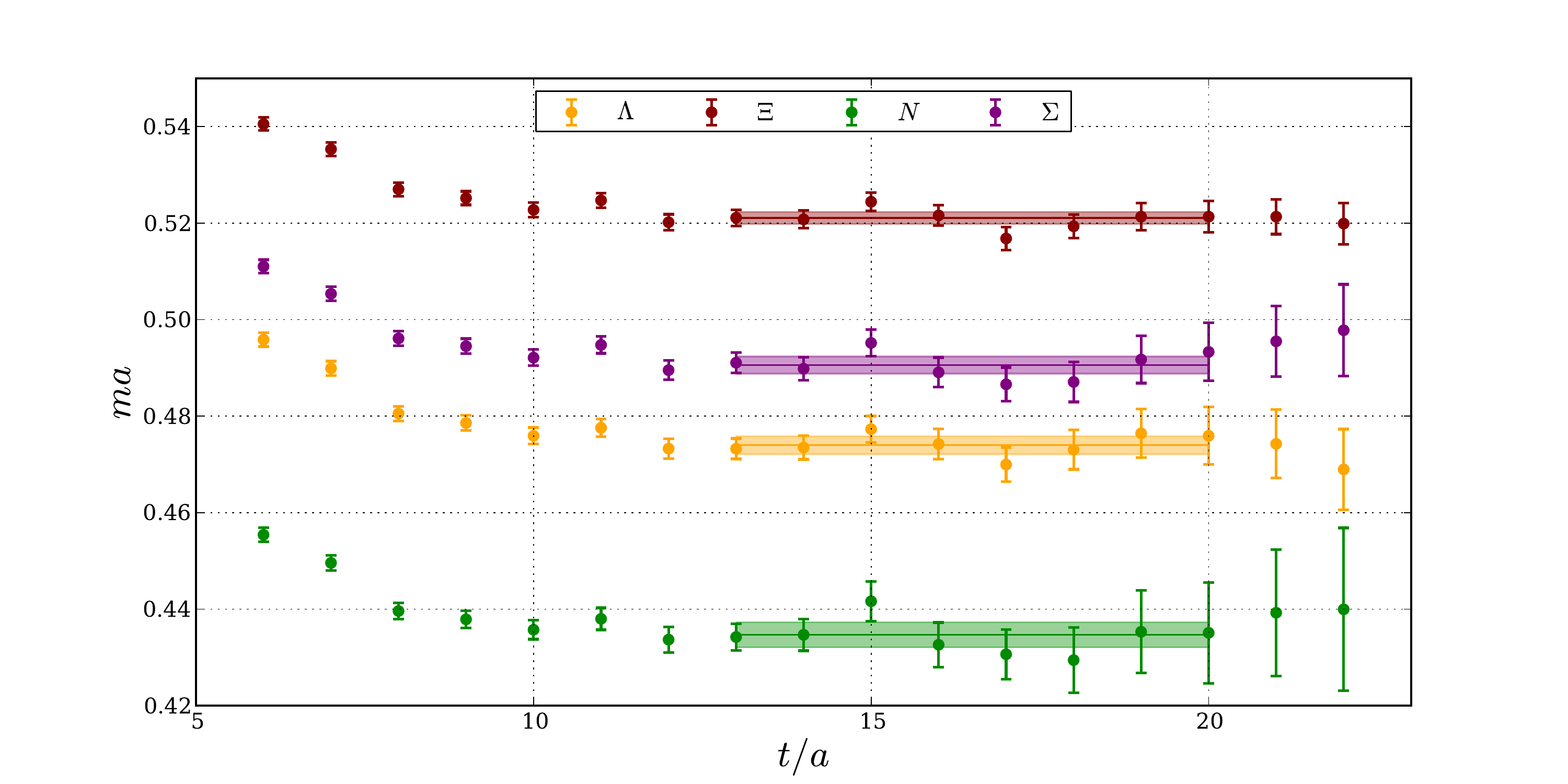}
\caption{The effective mass of the single-baryon correlation functions.}
\label{fig:efm_single}
\end{center}
\end{figure}

\begin{figure}[t!]
\includegraphics[width=0.5\linewidth]{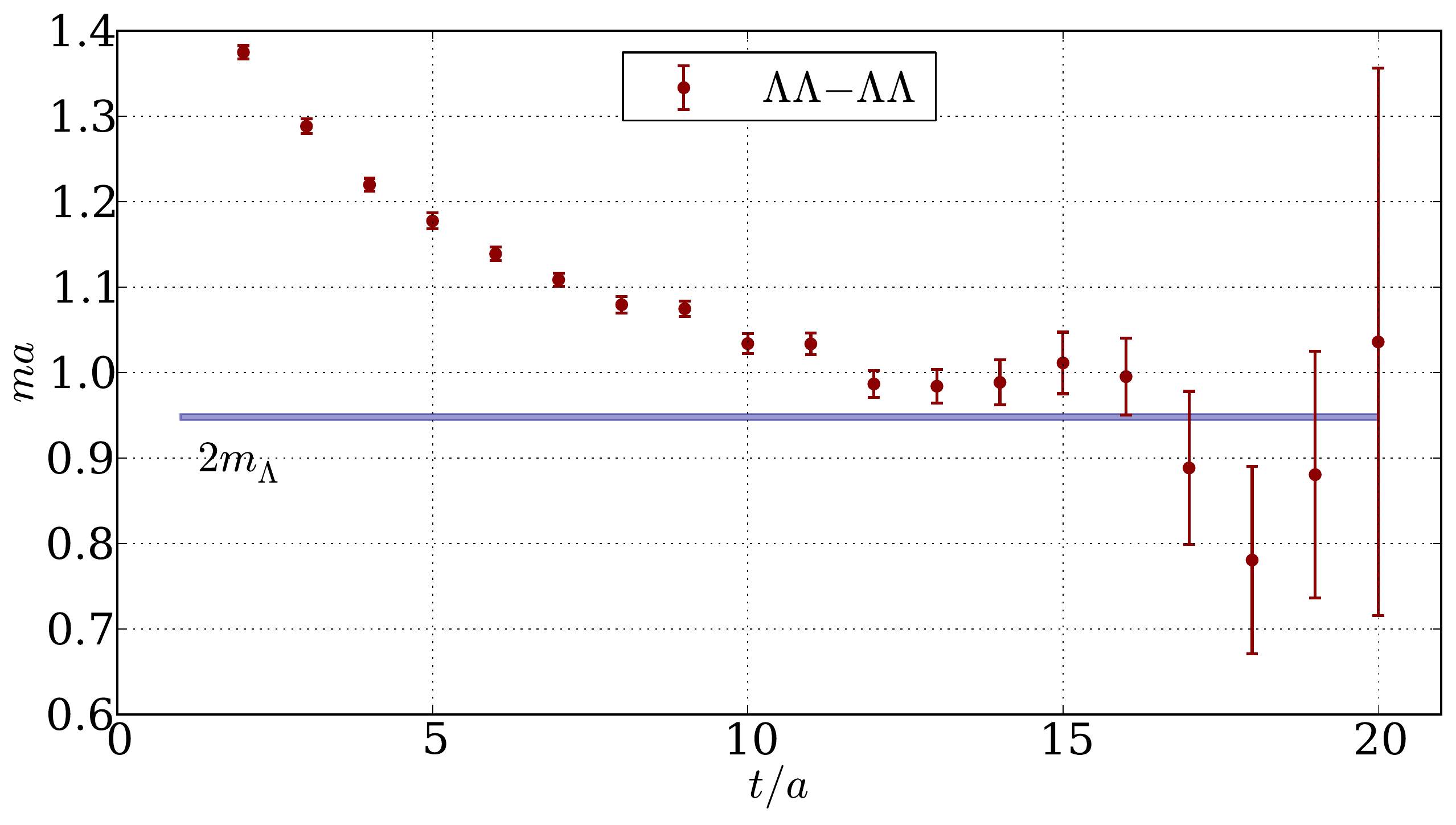}
\includegraphics[width=0.5\linewidth]{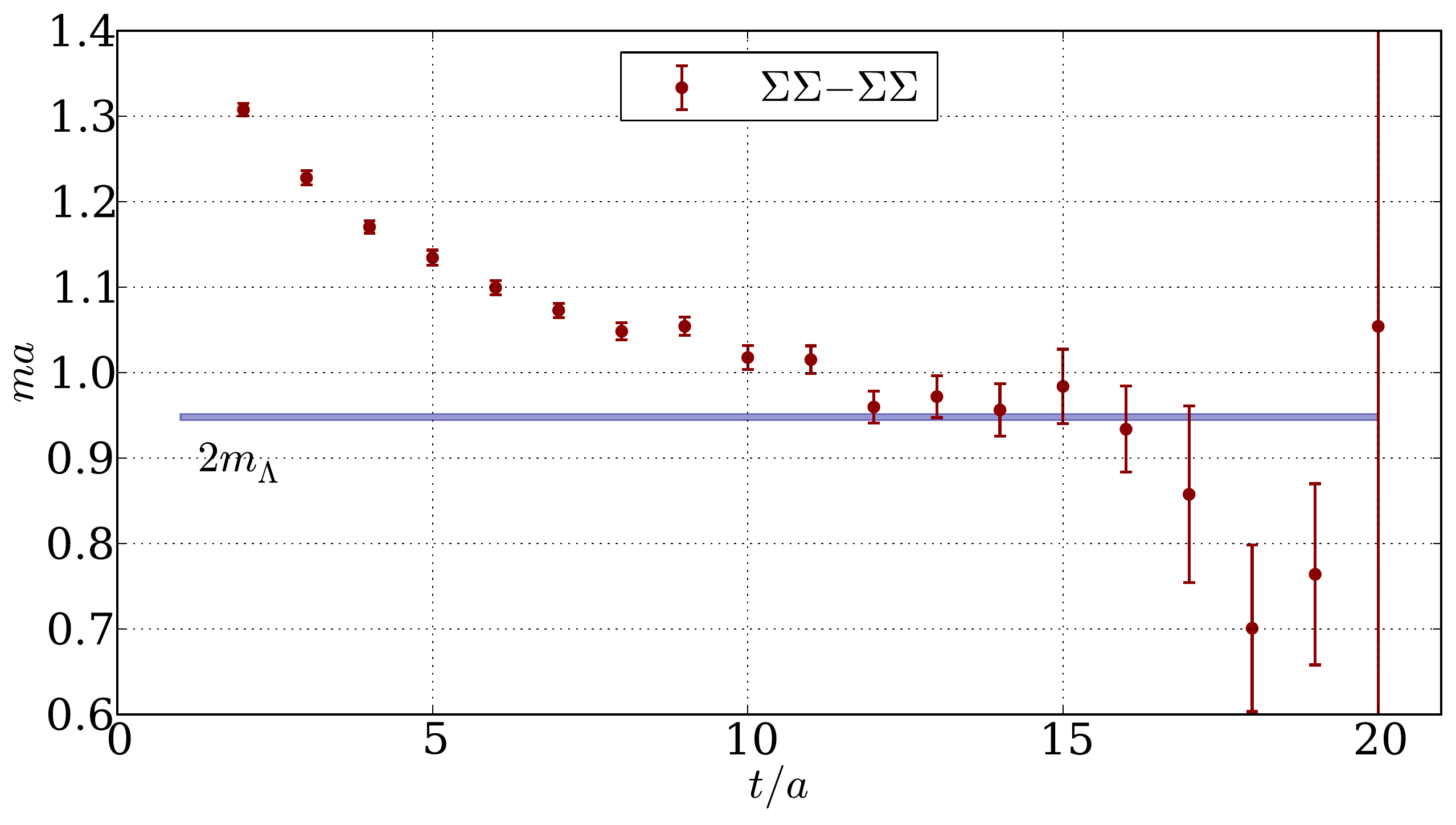}
\includegraphics[width=0.5\linewidth]{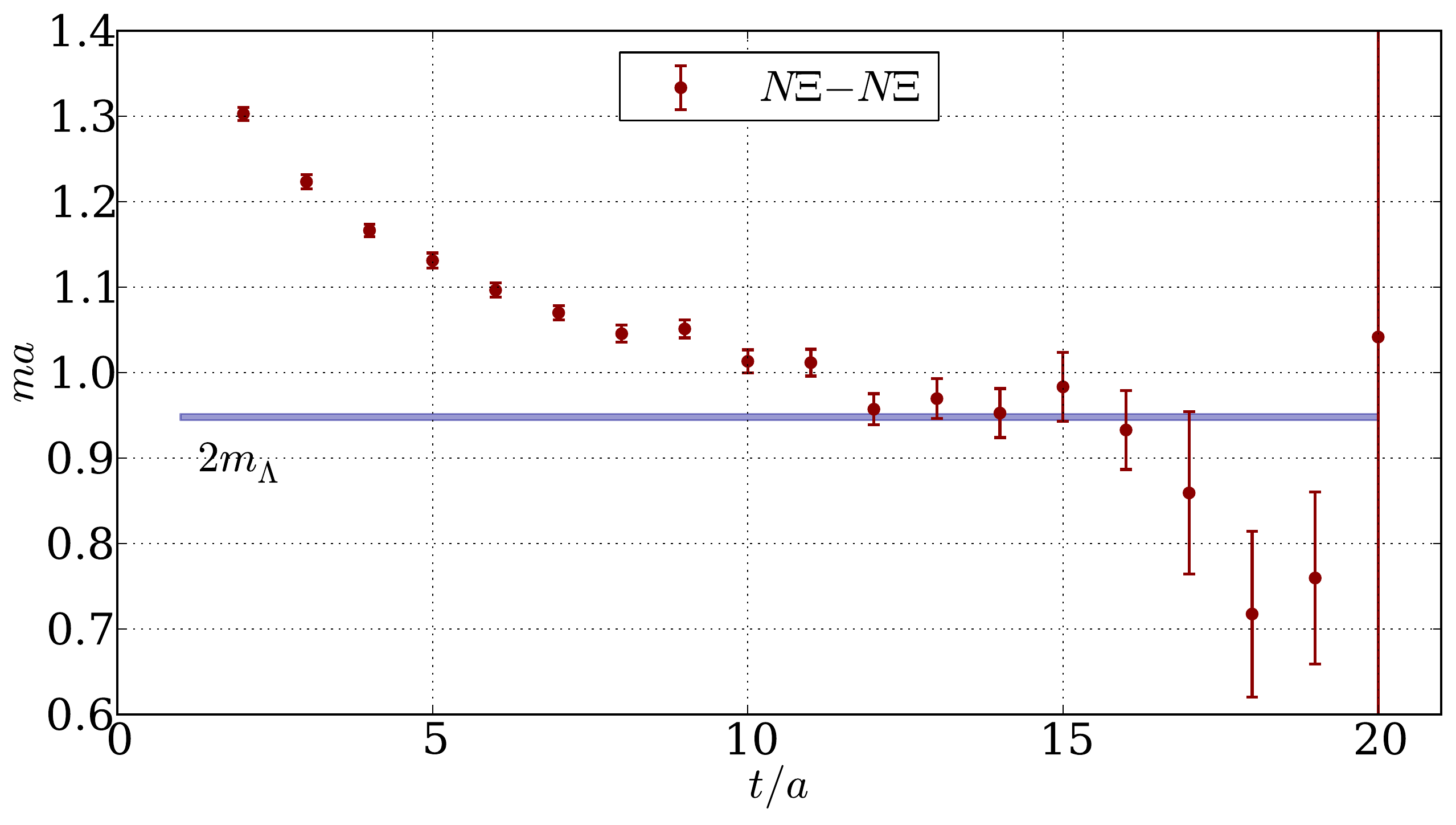}
\includegraphics[width=0.5\linewidth]{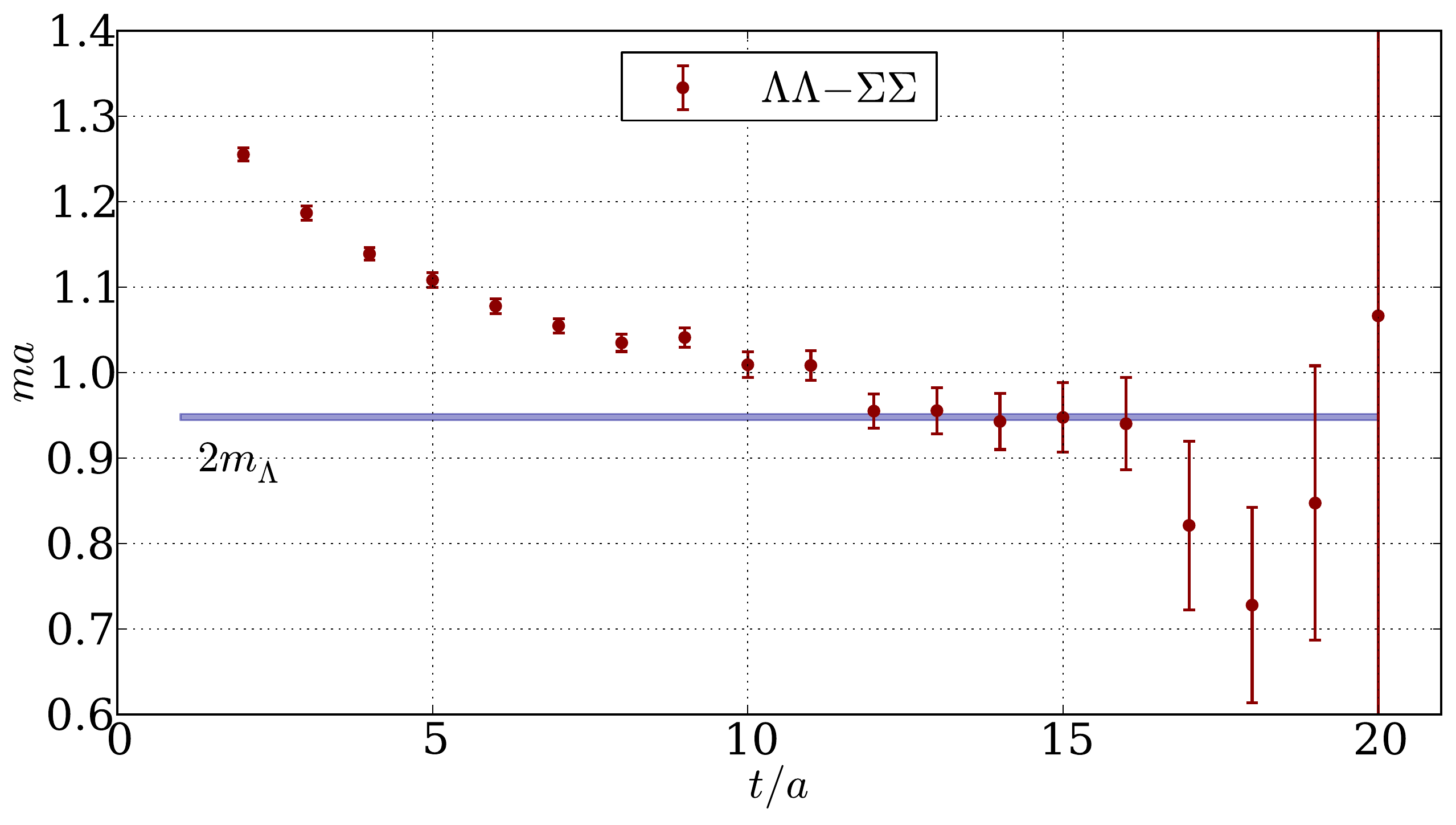}
\includegraphics[width=0.5\linewidth]{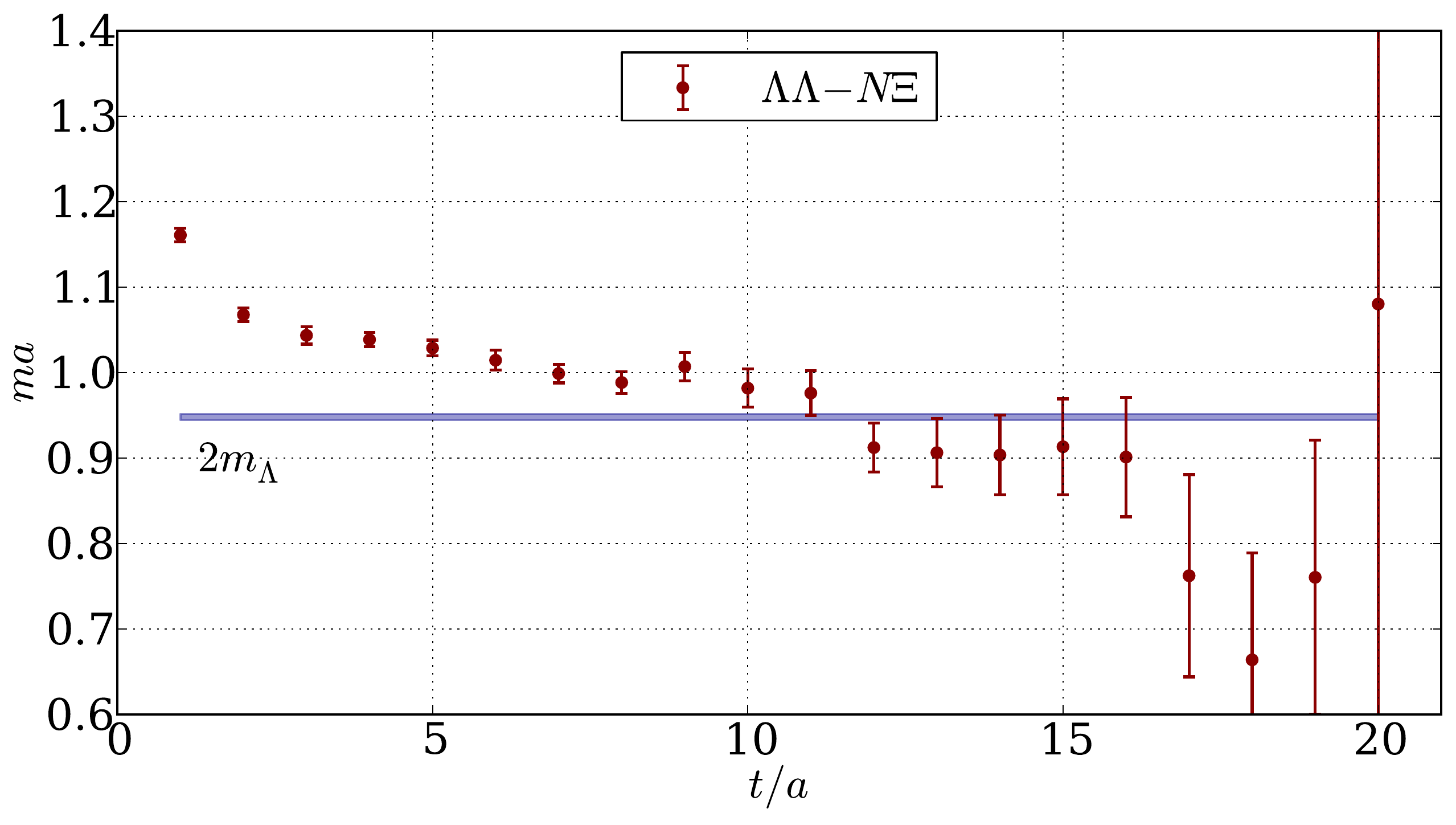}
\includegraphics[width=0.5\linewidth]{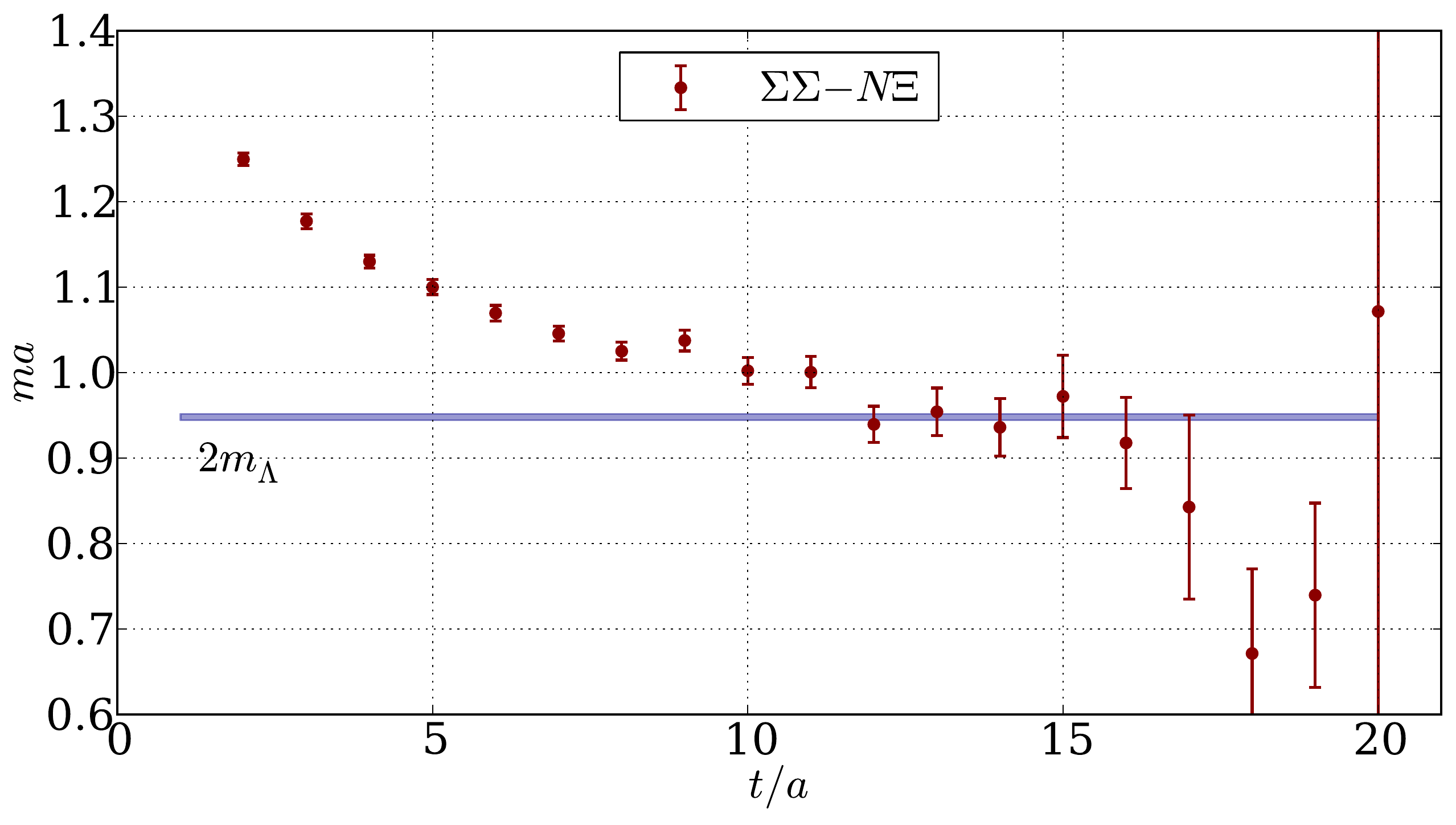}
\caption{The effective masses of the two-baryon correlation functions compared to twice the fitted value of the single $\Lambda-\Lambda$ system. From top left to bottom right: The $\Lambda\Lambda -\Lambda\Lambda$, $\Sigma\Sigma -\Sigma\Sigma$, $N\Xi -N\Xi$,
$\Lambda\Lambda -\Sigma\Sigma$, $\Lambda\Lambda-N\Xi$ and $\Sigma\Sigma -N\Xi$ channels.}
\label{fig:efm_dibaryon}
\end{figure}
\section{Numerical setup}
All our numerical results were obtained on dynamical gauge
configurations with two mass-degenerate quark flavours.  The gauge
action is the standard Wilson plaquette action \cite{Wilson:1974sk},
while the fermions were implemented via the O($a$) improved Wilson
discretization with non-perturbatively determined clover coefficient
$c_{\rm sw}$ \cite{Jansen:1998mx}.  The configurations
were generated within the CLS effort and the algorithms used are
based on L\"uscher's DD-HMC package~\cite{CLScode}.  

The correlation functions were calculated on a $64\times 32^3$ lattice (labeled `E5'
in~\cite{CLS}) with a lattice spacing of
$a=0.063$~fm and a pion mass of
$m_\pi=451$~MeV, so that $m_\pi L = 4.7$~\cite{Fritzsch:2012wq}. 
In addition we calculated correlation functions on the same ensembles with the mass parameter $\kappa$ tuned to produce a quark mass
comparable to that of the strange $m_q\simeq m_s^{phys}$~\cite{Capitani:2011fg}. 
This makes it possible to study two-baryon correlation functions away from the SU(3)-flavor limit with a partially quenched strange quark.
To improve the overlap of the interpolating operators with the baryons, we use Gaussian smearing \cite{smear:Gaussian89},
supplemented by APE smeared links \cite{Albanese:1987ds}, at both source and sink. The lattice ensemble consists of 900 independent gauge configurations, each separated by 16 trajectories. To boost the available statistics every configuration was inverted with a total of 16 source positions placed at a maximal separation within the lattice four-volume. In this way our full statistics amounts to 14400 measurements of the two-baryon correlation functions. 

\begin{figure}
\begin{center}
\includegraphics[width=.85\linewidth]{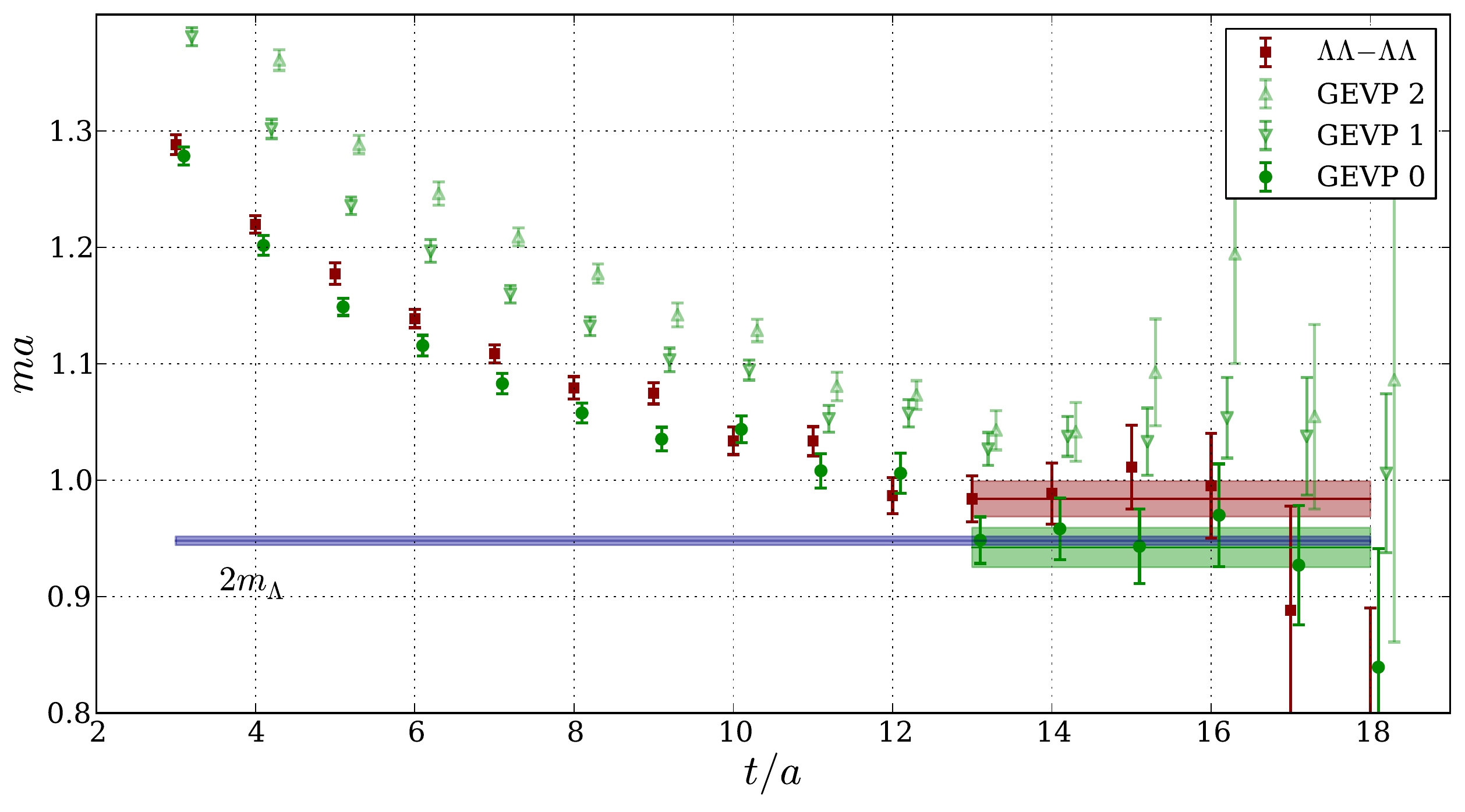}
\caption{GEVP showing the extracted groundstate (GEVP 0) and two excited states (GEVP 1 and 2), as well as the $\Lambda\Lambda-\Lambda\Lambda$ channel and the error band of twice the single $\Lambda-\Lambda$ for reference.}
\label{fig:GEVP}
\end{center}
\end{figure}

\section{Results on single-baryon and two-baryon correlation functions}

Ultimately, the goal is to determine whether two-baryon correlation functions allow for a state with a mass below that of two individual (unbound) $\Lambda$ particles. Therefore we study the single-baryon spectrum in addition to the more complicated two-baryon case using the effective mass

\begin{equation}
ma\ (t) = \frac{1}{t_J} \log \frac{C(t)}{C(t+t_J)}\ ,
\end{equation}
\newline
where we choose $t_J = 3$ (this improves the stability of the effective mass plot over $t_J = 1$, 
as neighbouring points are more correlated than those with a larger separation).
We show the effective mass for the $\Lambda-\Lambda$, $\Xi-\Xi$, $\Sigma-\Sigma$ and $N-N$ correlation functions in Fig.~\ref{fig:efm_single}. As expected, the effective mass plot exponentially decreases before forming a plateau at large times $t/a\simeq10$, from which the mass of the ground state can be extracted. 
Clearly the large number of measurements yields stable plateaus with errors on the individual points at the 1\%-level for the nucleon and sub-1\%-level for the baryons containing a strange quark. This results in errors of $\sim0.5\%$ for the fitted values of the single baryon spectrum masses. This indicates the uncertainty of determining the bound or unbound nature of the dibaryons will be limited by the error of the two-baryon correlation function.
The effective mass results for all six dibaryon correlation functions of Eq.~\ref{eq:bbmatrix} are shown in Fig.~\ref{fig:efm_dibaryon}. For comparison, we give the error band of twice the single $\Lambda-\Lambda$ as a blue line in all figures. A clear undercutting of this line by the two-baryon results would immediately indicate a bound dibaryon state in the corresponding correlator channel. At the current level of statistics our analysis does not clearly indicate whether these dibaryon channels are bound or not. 
\begin{figure}[t!]
\includegraphics[width=0.5\linewidth]{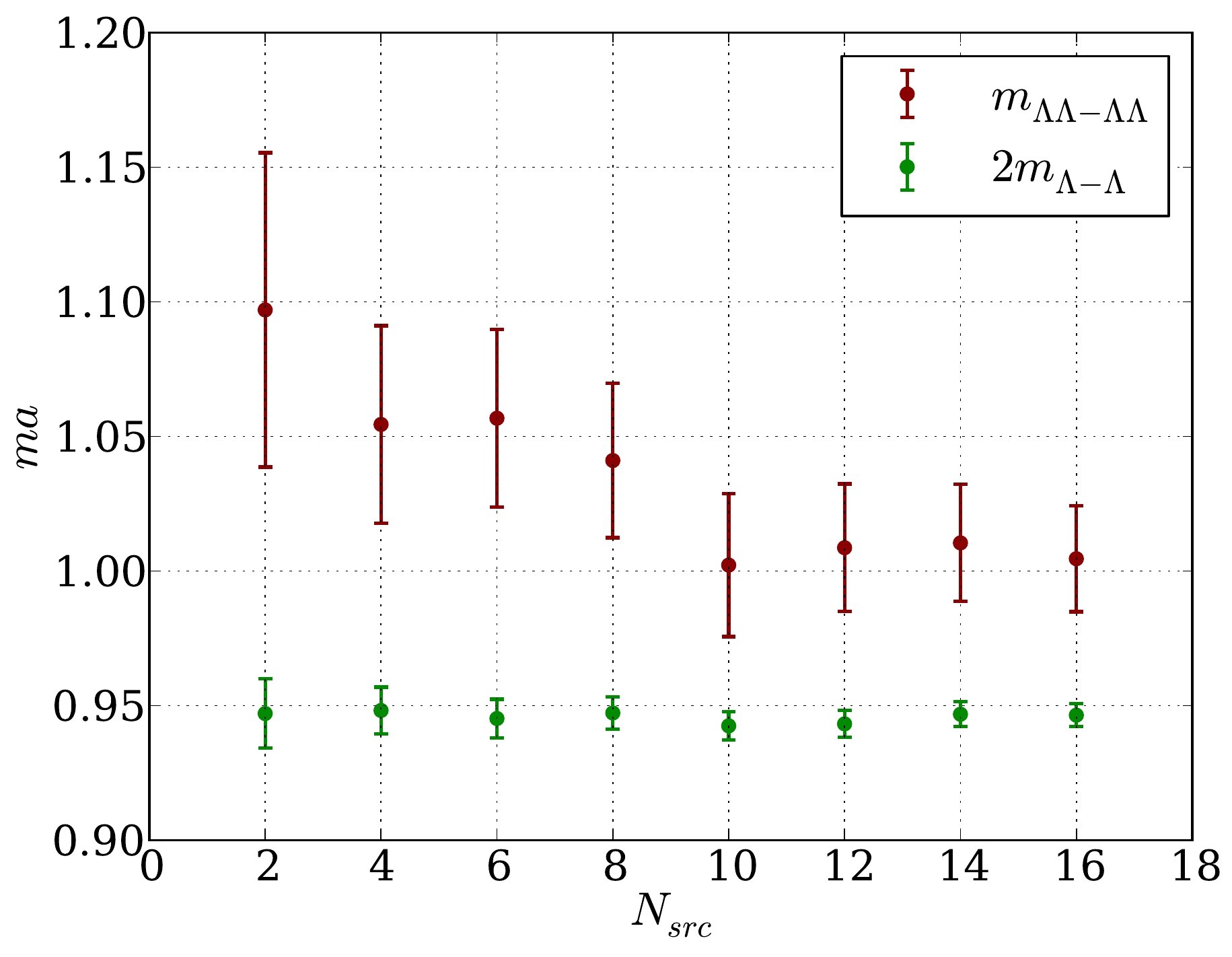}
\includegraphics[width=0.5\linewidth]{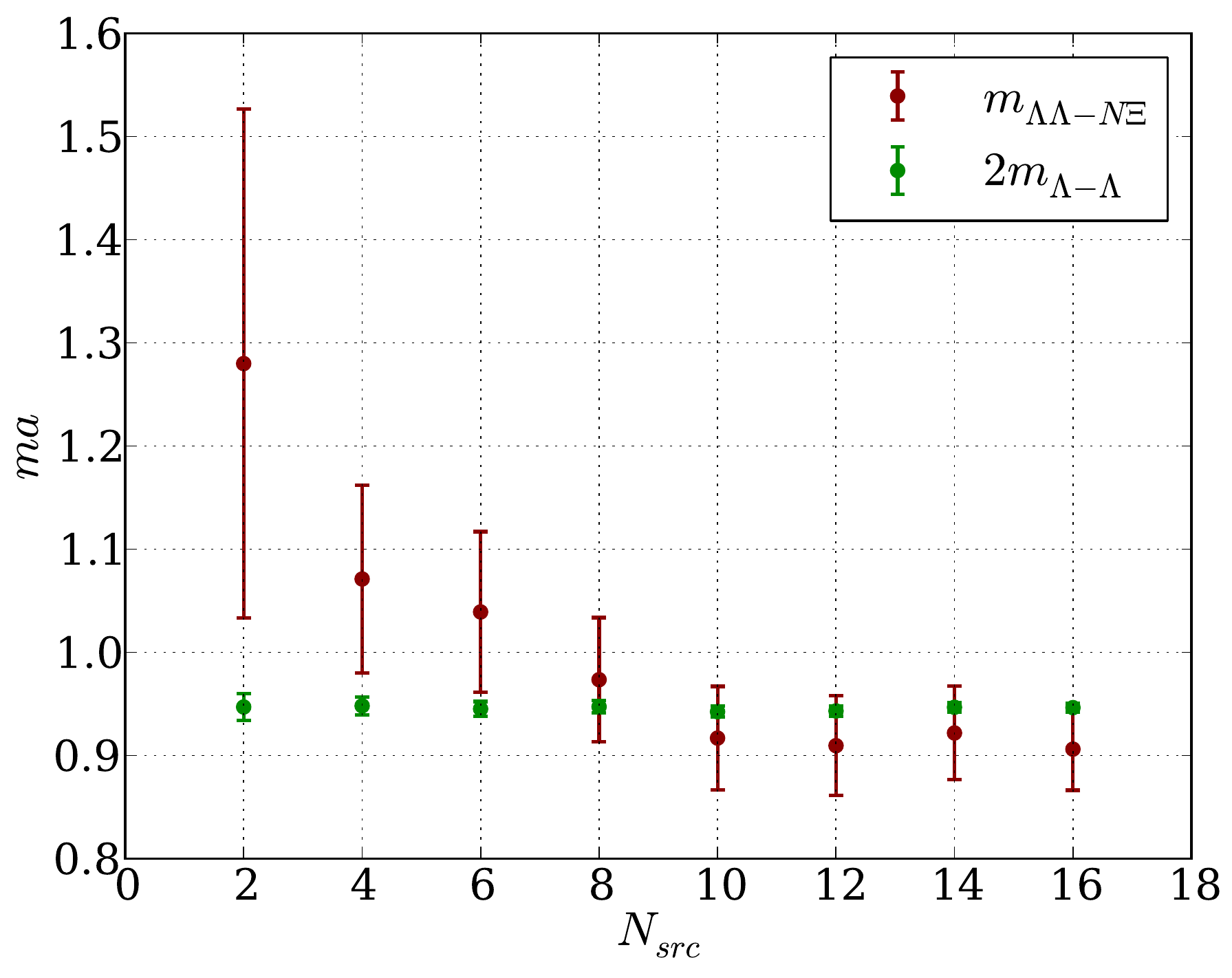}
\caption{The effective mass at $t/a=13$ of twice the $\Lambda-\Lambda$ single-baryon (green) and the $\Lambda\Lambda-\Lambda\Lambda$ (left) or $\Lambda\Lambda -N\Xi$ (right) two-baryon correlators over the number of sources per configuration. The error is seen to scale as $\sqrt{N_{src}}$ as expected. Note: All results are given in lattice units.}
\label{fig:efm_sources}
\end{figure}
We note that
compared to the single-baryon case the approach to the asymptotic behaviour is considerably slower. This may be explained by the neighbouring states in the correlation function being closer together in the dibaryon than in the single-baryon cases. As a consequence states with a slightly larger mass take more time to decay, which makes a definitive determination of the ground state difficult. The effect is especially visible in the unmixed channels, such as the $\Lambda\Lambda-\Lambda\Lambda$. In the mixed case, such as the $\Lambda\Lambda-N \Xi$, the plateau appears to be reached earlier. 
However, unlike the unmixed channels, the matrix elements in the mixed channels need not be positive and so there is the possibility
of cancellations between different states. It is therefore important to take into account the mixing between different states. A natural way to do this is using the six correlation functions as input into a GEVP \cite{GEVP1,GEVP2,GEVP3}, which we show preliminary results for in Fig.~\ref{fig:GEVP}. 
This demonstrates an improvement in the extraction of the ground state over studying a single channel such as the $\Lambda\Lambda-\Lambda\Lambda$. Within current statistics it is difficult to differentiate whether this state is weakly bound or unbound and therefore a further boost in statistics is required. We will extend the GEVP analysis in a forthcoming publication through the use of multiple smearing operators. Currently, we have only used smeared-smeared operators, adding both smeared-local and local-local operators to the analysis will extend the matrix Eq.~\ref{eq:bbmatrix} from $3\times3$ to $6\times6$ entries, which should improve the effectiveness of the GEVP and thus the resolution of the ground state. 
\newline\indent To check the dependence of our result on the statistics and number of source positions used in the calculation, we show the results of the effective mass at $t/a=13$ in the single $\Lambda-\Lambda$ and the two-baryon $\Lambda\Lambda-\Lambda\Lambda$ (left) or
$\Lambda\Lambda-N\Xi$ (right) channels in Fig.~\ref{fig:efm_sources}. The error decreases as the number of sources is increased from roughly $\sigma_{2}\simeq0.060$ to $\sigma_{16}\simeq0.022$ in the $\Lambda\Lambda-\Lambda\Lambda$ channel. This decrease in the statistical error is given by $\sigma(N_{src})\simeq\sqrt{N_{src}}$, as expected. In the dibaryon channels the central value of the effective mass at $t/a=13$ also decreases with increasing $N_\textrm{src}$, while the single-baryon value remains almost constant. This further indicates that a higher statistical accuracy is required for a conclusive result on the bound or unbound nature of the dibaryons. 

\section{Conclusion}
Researching multi-baryon systems such as the H-dibaryon using lattice QCD methods is a challenging problem, as it poses large demands of the algorithms used and the number of statistics required. Here, we have presented a preliminary study of dibaryon correlation functions (that have an overlap with the H-dibaryon) using O($a$) improved Wilson fermions and $N_f=2$ dynamical gauge configurations made available to us through the CLS effort \cite{CLS}. We have implemented a blocking algorithm in the spirit of \cite{Endres,Det} that handles the required contractions and that can be easily extended to correlation functions of the larger mass number nuclei. The string of configurations used in this study is long, boasting 900 independent configurations. Nevertheless, the study of the hyperon-hyperon correlation functions require the use of inversions at multiple source positions. We tested the impact of the number of sources on our calculation and found (a) the $\sqrt{N_{src}}$ error reduction behaviour is indeed observed and (b) still more statistics will be needed to decide on whether the dibaryon ground state is bound or not in our calculation. We have demonstrated the advantage of using a GEVP analysis, which coupled with improved statistics will better
determine the nature of the hyperon-hyperon correlation function and ultimately the H-dibaryon.

~\newline\noindent
{\bf Acknowledgments:}
We thank Georg von Hippel and Harvey Meyer for inspiring discussions, and to our colleagues within CLS for sharing the lattice ensemble used. All correlators were computed on the dedicated QCD platform "Wilson" at the Institute for Nuclear Physics, University of Mainz.

\end{document}